\documentstyle[aps,twocolumn]{revtex}

\begin{document}
\draft
\title{\noindent
Proposal for preparing delocalised mesoscopic states of
material oscillators
}

\author{A.S. Parkins}
\address{
Department of Physics, University of Auckland, Private Bag 92019, 
Auckland, New Zealand
}

\maketitle

\begin{abstract}
A scheme is proposed for preparing delocalised mesoscopic
states of the motion of two or more atoms trapped at distantly-separated 
locations. Generation of entanglement is achieved using 
interactions in cavity quantum electrodynamics which 
facilitate motional quantum state transmission, via light, 
between separate nodes of a quantum network. Possible applications
of the scheme are discussed.
\end{abstract}


\narrowtext

\section{
Introduction
}

The superposition principle and nonlocality are aspects of
quantum mechanics that have teased and enticed the scientific
and general communities alike since its development earlier this 
century. 
More specifically, conceptual difficulties arise when one tries
to reconcile these features with the behaviour of the 
macroscopic, everyday world with which we are most familiar.
These difficulties are exemplified 
by the classic Schr\"odinger Cat \cite{Schrodinger35} and
Einstein-Podolsky-Rosen (EPR) \cite{Einstein35} paradoxes.

The boundary between the quantum and classical worlds remains
a very active area of research, with, in particular,
the atomic physics and quantum optics communities recently 
providing a number of significant experimental demonstrations.
These include the preparation of mesoscopic superposition 
states -- in particular, superpositions of small-amplitude 
coherent states -- of a material oscillator (a trapped atom) 
\cite{Monroe96} and of a cavity radiation field \cite{Brune96},
and the generation of quantum-mechanically entangled internal
states of pairs \cite{Hagley97,Turchette98,Rauschenbeutel00} 
and quadruplets \cite{Sackett00} of atoms. 
These experiments open the door
to detailed and highly-controlled investigations of quantum
measurement, quantum decoherence, and nonlocality in mesoscopic 
systems and with massive particles.

Further along these lines, in this work we consider a 
combination of the technologies used in the above demonstrations,
that is, single-atom trapping and cavity quantum electrodynamics
(QED), to propose a scheme for the preparation of, arguably, 
a yet more exotic quantum state -- a
delocalised mesoscopic superposition state of two or more 
distantly-separated trapped atoms; that is, a state of
the form
\begin{eqnarray} \label{state}
&& \frac{1}{\sqrt{\cal N}} \left( \,
|\alpha\rangle^1 |0\rangle^2 \cdots |0\rangle^N +
|0\rangle^1 |\alpha\rangle^2 \cdots |0\rangle^N + \ldots \right.
\nonumber
\\
&& \;\;\;\;\;\;\;\;\;\;\;\;\;\;\; \left. + \,
|0\rangle^1 |0\rangle^2 \cdots |\alpha\rangle^N \, \right) \, ,
\end{eqnarray}
where $|\alpha\rangle^j$ denotes a coherent state (in one dimension) 
of the {\em motion of the atom} at the $j$-th site, and 
${\cal N}$ is a normalisation factor.
The term `mesoscopic' is used because the 
coherent state $|\alpha\rangle^j$ is a `classical-like' state,
for which the mean number of quanta, $\bar{n}=|\alpha |^2$, may 
be relatively large (within the constraints of our scheme). 

Before continuing, 
we make note of proposals that exist for the preparation of 
separated {\em light fields} in states of the form (\ref{state})
with $N=2$ (see, for example, \cite{Sanders92,Davidovich93}).
Again though, it should be emphasised that 
here we are dealing with the states of {\em massive particles}.

\section{
Quantum state exchange between motion and light
}

Central to the proposed scheme is the ability to exchange quantum
states between the motion of a trapped atom and a light field,
allowing long-distance transmission of quantum information.
As described in recent work \cite{Parkins99}, 
this can be done
using a set of interactions in cavity QED, which we now
summarise.

\subsection{
Trapped atom coupled to an optical cavity mode
}

The basic setup used
was originally considered by Zeng and Lin \cite{Zeng94}. 
This setup consists of a two-level atom (or ion) confined
in a harmonic trap located inside an optical cavity. 
The atomic transition of frequency
$\omega_0$ is coupled to a single mode of the cavity field of
frequency $\omega_{\rm cav}$ and is also driven by an
external (classical) laser field of frequency $\omega_{\rm A}$.
The physical setup and excitation scheme are
depicted in Fig.~1,
with the relevant internal atomic levels being $|\uparrow\rangle$
and $|e\rangle$. 
The cavity is aligned along the $x$-axis, while
the laser field [laser A in Fig.~1(b)] is incident from a direction 
in the $y$-$z$ plane.
The additional lasers B,C, and D and internal atomic levels 
$|\downarrow\rangle$ and $|f\rangle$ will be discussed later, but
can be ignored for the moment.

The Hamiltonian describing the atom-cavity system takes the  
form (in a frame rotating at the laser frequency $\omega_{\rm A}$)
\begin{eqnarray} \label{H}
\hat{H} = && 
\sum_{j=x,y,z} 
\hbar\nu_j (\hat{b}_j^\dagger\hat{b}_j+1/2) + \hbar\delta
\hat{a}^\dagger\hat{a} + \hbar\Delta\hat{\sigma}_+\hat{\sigma}_-
\nonumber
\\
&& \;\;\; +\, \hbar 
\left[ {\cal E}_{\rm A}(\hat{y},\hat{z},t) 
\hat{\sigma}_+ + 
{\cal E}_{\rm A}^\ast (\hat{y},\hat{z},t)
\hat{\sigma}_- \right] \nonumber
\\
&& \;\;\; +\, \hbar
g_0 \sin (k\hat{x}) (\hat{a}^\dagger
\hat{\sigma}_- + \hat{\sigma}_+\hat{a} )  \nonumber
\\
&& \;\;\; +\, 
\hat{a}^\dagger \hat{\Upsilon}_{\rm c} + 
\hat{\Upsilon}_{\rm c}^\dagger \hat{a} +\, 
\hat{\sigma}_+ \hat{\Upsilon}_{\rm a} + 
\hat{\Upsilon}_{\rm a}^\dagger \hat{\sigma}_- \, .
\end{eqnarray}
Here, $\{\nu_x,\nu_y,\nu_z\}$ are the harmonic oscillation 
frequencies along the principal axes of the trap, 
$\hat{b}_j$ and $\hat{a}$ are annihilation operators for the
quantised atomic motion and cavity field, respectively, 
$\hat{\sigma}_-=|\uparrow\rangle\langle e|$ is the atomic lowering 
operator for the $|\uparrow\rangle\leftrightarrow |e\rangle$ transition, 
and $\delta =\omega_{\rm cav}-\omega_{\rm A}$ and 
$\Delta =\omega_0-\omega_{\rm A}$. 
The quantity ${\cal E}_{\rm A}(\hat{y},\hat{z},t)$ is the
(time-dependent) amplitude of laser field A.
The single-photon atom-cavity dipole coupling strength is
given by $g_0$, while the sine function describes the standing wave
structure of the cavity field (we assume that the centre of the 
trap is located at a {\em node} of the cavity field), with 
$k=2\pi /\lambda$ the wavenumber of the field and 
$\hat{x}=[\hbar /(2m\nu_x)]^{1/2}(\hat{b}_x+\hat{b}_x^\dagger )$. 
Finally, the last line in (\ref{H}) describes the coupling of
the internal cavity field mode to the `reservoir' of external 
field modes (with $\hat{\Upsilon}_{\rm c}$ the `reservoir annihilation
operator'), which produces damping of the cavity field, and
the coupling of the atomic transition to (vacuum)
radiation field modes other than the cavity mode, which gives rise
to free-space spontaneous emission \cite{Walls94}.
Note that we neglect any forms of motional decoherence associated 
with the trap itself.

In \cite{Parkins99} a number of assumptions and approximations are
made in order to simplify the model. In particular:
\begin{enumerate}
\item
The detunings of the light fields from the atomic transition 
frequency are assumed to be very large 
[that is, $\Delta\gg |{\cal E}_{\rm A}(t)|,g_0,\delta,\nu_j$], enabling 
atomic spontaneous emission to be neglected and the internal atomic
dynamics to be adiabatically eliminated.
\item
The size of the harmonic trap is assumed to be small compared to the
optical wavelength (Lamb-Dicke regime), enabling the approximations
$\sin (k\hat{x})\simeq \eta_x (\hat{b}_x+\hat{b}_x^\dagger )$ and 
${\cal E}_{\rm A}(\hat{y},\hat{z},t)\simeq
{\cal E}_{\rm A}(t){\rm e}^{-i\phi_{\rm A}}$, 
where
$\eta_x$ is the Lamb-Dicke parameter for the $x$-axis.
Note that this assumption ultimately places a restriction on the
mean excitation numbers $\{\bar{n}_j\}$ associated with the vibrational
state of the atom, since the atomic wavepacket broadens with 
increasing $\bar{n}_j$. We will return to this point in the discussion 
when we consider the experimental feasibility of the scheme
\item
The cavity and laser fields are tuned so that 
$\delta =\omega_{\rm cav}-\omega_{\rm A}=\nu_x$; that is, the fields
are tuned to drive Raman transitions between neighbouring vibrational
levels [see Fig.~1(b)].
\item
The trap frequency $\nu_x$ and cavity field decay rate $\kappa$
are assumed to satisfy 
$\nu_x\gg\kappa\gg |(g_0\eta_x/\Delta ){\cal E}_{\rm A}(t)|$.
The first inequality allows a rotating-wave approximation to be
made with respect to the trap oscillation frequency, while the
second inequality enables an adiabatic elimination of the cavity
field mode.
\end{enumerate}
Given these conditions one can show that the description of the
motional mode dynamics in the $x$ direction can be reduced to the simple 
quantum Langevin equation
\begin{equation} \label{QLE}
\dot{\hat{b}}_x \simeq 
- [\Gamma (t)+i\nu_x] \hat{b}_x + e^{i\phi_{\rm A}} \sqrt{2\Gamma (t)} \,
e^{-i\nu_xt} \hat{a}_{\rm in}(t)\; ,
\end{equation}
where
\begin{equation}
\Gamma (t) = \frac{1}{\kappa}\, \left[
\frac{g_0\eta_x{\cal E}_{\rm A}(t)}{\Delta} \right]^2 .
\end{equation}
The operator $\hat{a}_{\rm in}(t)$ obeys the commutation
relation 
$[\hat{a}_{\rm in}(t),\hat{a}_{\rm in}^\dagger (t^\prime )]
=\delta (t-t^\prime )$ and describes the quantum noise input to
the cavity field (in a frame rotating at the cavity frequency)
through the partially transmitting input/output mirror [see Fig.~1(a)].

\subsection{
Motional-state transfer between distant sites
}

The significance of Eq.(\ref{QLE}) is that it amounts to a
simple coupling of the motional mode to propagating
light modes external to the cavity. 
More precisely, from the input-output theory of optical
cavities \cite{Walls94,Gardiner00}, it can be shown that the cavity 
output field is given, under the present circumstances, by
\begin{equation} \label{aout}
\hat{a}_{\rm out}(t) \simeq - \, \hat{a}_{\rm in}(t) -
\sqrt{2\Gamma (t)} \; e^{i\nu_xt} \hat{b}_x(t) \, ,
\end{equation}
where we have set $\phi_{\rm A}=0$ for simplicity.
Following work by Cirac {\em et al}. \cite{Cirac97} (see also
\cite{Gardiner00}) on the
transmission of a qubit between two nodes of a quantum network, 
it is shown in \cite{Parkins99} that if the output field from one
of our atom-cavity configurations is incident on a second such
atom-cavity configuration, with the coupling between systems being
{\em unidirectional}, then with suitably tailored laser pulses 
${\cal E}_{\rm A1}(t)$ and ${\cal E}_{\rm A2}(t)$ applied at the
two sites one may realise the motional state transfer
(assuming both atoms to be in the internal state $|\uparrow\rangle$)
\begin{equation} \label{statetransfer}
|\phi\rangle_x^1 \, |0\rangle_x^2 \rightarrow
|0\rangle_x^1 \, |\phi\rangle_x^2 \, ,
\end{equation}
where $|\phi\rangle_x$ is an {\em arbitrary} quantum state
describing the motion along the $x$-axis.
Such a state transfer configuration is illustrated
schematically in Fig.~2, while example 
pulse shapes, specified through the effective coupling rates 
of the motional modes to the external light fields,
$\Gamma_1(t)$ and $\Gamma_2(t)$, are \cite{Parkins99}
\begin{equation}
\Gamma_1(t) = \Gamma \, \frac{{\rm e}^{\Gamma t}}
{{\rm e}^{\Gamma t}+{\rm e}^{-\Gamma t}} \, , \;\;\;\;
\Gamma_2(t) = \Gamma_1(-t) \, ,
\end{equation}
assuming the transfer starts at $t=-\infty$ and concludes at
$t=+\infty$, with $\Gamma$ a constant 
[equal to the maximum value of $\Gamma_{1,2}(t)]$ \cite{PulseShapes}.
Armed with this capability, we are able to distribute quantum
states of a material oscillator, and generate entanglement,
between macroscopically-separated locations, leading to 
some fascinating possibilities, one of which we now consider.

\section{
Preparation of a delocalised motional coherent state
}

In addition to the atom-cavity coupling, our scheme for the
preparation of a delocalised coherent state  
utilises auxiliary lasers B,C, and D and internal atomic levels
$|\downarrow\rangle$ and $|f\rangle$.
Lasers B and C induce coherent Raman transitions between the
internal states $|\downarrow\rangle$ and $|\uparrow\rangle$, and 
their frequency difference is chosen to be resonant with the
transition frequency $\omega_\uparrow -\omega_\downarrow$, so that
the motional state of the trapped atom is unaffected \cite{Monroe96}. 
It is further assumed that if the atom is in the internal state
$|\downarrow\rangle$ then it does not `see' the cavity field or
the coupling laser A. Hence, the motional state transfer described
above only occurs when the atoms are in the internal state 
$|\uparrow\rangle$.
Laser field D, when switched on, resonantly excites the transition
$|\downarrow\rangle\leftrightarrow |f\rangle$; detection (absence) 
of fluorescence from this transition projects the system onto the
internal state $|\downarrow\rangle$ ($|\uparrow\rangle$)
\cite{Monroe96,Wineland98a}.
Finally, while we focus on the (quasi-classical) coherent state 
$|\alpha\rangle_x$, it should be noted that the scheme outlined below
will work for an arbitrary motional state $|\phi\rangle_x$.

\subsection{
Two atoms
}

Consider first the configuration shown in Fig.~2. Atom 1 is
assumed to be prepared, by some means (for example, using additional 
laser or electric fields \cite{Monroe96,Wineland98a}), 
in a coherent state of
its motion along the $x$-axis, while its internal state is
prepared, via lasers B1 and C1, as the superposition
$2^{-1/2}(\, |\uparrow\rangle^1 +|\downarrow\rangle^1 \, )$.
Atom 2 is prepared in its ground motional state 
$|0\rangle_x^2$ and in the
internal state $|\uparrow\rangle^2$. Applying the state
transfer laser pulses (through lasers A1 and A2 at sites 1 and 2,
respectively) produces the transformation
\begin{eqnarray} \label{eq1}
\frac{1}{\sqrt{2}} && \left( \, |\uparrow\rangle^1 +
|\downarrow\rangle^1 \, \right) |\alpha\rangle_x^1 \,
|\uparrow\rangle^2 |0\rangle_x^2 \nonumber
\\
&& \rightarrow \frac{1}{\sqrt{2}} \left( \,
|\uparrow\rangle^1 |0\rangle_x^1 |\alpha\rangle_x^2
+ |\downarrow\rangle^1 |\alpha\rangle_x^1 |0\rangle_x^2
\, \right) |\uparrow\rangle^2 \, .
\end{eqnarray}
Importantly, the part involving $|\downarrow\rangle^1$
does not change, as, again, from this state atom 1 does not couple
to the laser field A1 or to the cavity field (so, cavity 1 is
not excited and no light propagates to cavity 2, which also 
means that atom 2 remains in its ground motional state).
Next, at site 1, lasers B1 and C1 are used to produce the
internal state transformations
\begin{eqnarray}
|\uparrow\rangle^1 &\rightarrow & \frac{1}{\sqrt{2}} \left( \,
|\uparrow\rangle^1 - |\downarrow\rangle^1 \, \right) \, ,
\nonumber
\\
|\downarrow\rangle^1 &\rightarrow & \frac{1}{\sqrt{2}} \left( \,
|\uparrow\rangle^1 + |\downarrow\rangle^1 \, \right) \, ,
\end{eqnarray}
leading to an overall system state
\begin{eqnarray}
&&  \frac{1}{2} \left[ \,
|\uparrow\rangle^1 \left( \, |0\rangle_x^1 |\alpha\rangle_x^2
+ |\alpha\rangle_x^1 |0\rangle_x^2 \, \right) \right.
\nonumber
\\
&& \;\;\;\;\;\;\;\; \left. +\, |\downarrow\rangle^1 \left( \, 
-|0\rangle_x^1 |\alpha\rangle_x^2 + |\alpha\rangle_x^1 
|0\rangle_x^2 \, \right) \, \right] \, |\uparrow\rangle^2 \, .
\end{eqnarray}
Correlated with each internal state of atom 1 is thus a 
motional coherent state delocalised between atoms 1 and 2.
If now laser D1 is applied to atom 1, then a null detection
of fluorescence (occurring 50\% of the time) will project the 
system into the state 
\begin{equation} \label{2atom}
\frac{1}{\sqrt{\cal N}} \left( \, |0\rangle_x^1 |\alpha\rangle_x^2
+ |\alpha\rangle_x^1 |0\rangle_x^2 \, \right) \, 
|\uparrow\rangle^1 |\uparrow\rangle^2 .
\end{equation}
Note that detection of fluorescence, or more particularly the
spontaneous scattering of photons associated with this fluorescence, 
would be expected to perturb the motional state in an uncontrollable 
and undesirable manner.

\subsection{
Three atoms
}

By adding another atom-cavity system to the cascade of Fig.~2,
it is straightforward to extend the above scheme to prepare
a motional coherent state delocalised between {\em three} atoms.
Taking now the initial state of the system to be
\begin{equation} \label{init3}
\frac{1}{\sqrt{5}} \left( \, 2|\uparrow\rangle^1 +
|\downarrow\rangle^1 \, \right) \, |\alpha\rangle_x^1 \,
|\uparrow\rangle^2 |0\rangle_x^2 \, 
|\uparrow\rangle^3 |0\rangle_x^3 \, ,
\end{equation}
applying the state transfer pulses (lasers A1 and A2) 
to atoms 1 and 2 transforms this state to
\begin{equation}
\frac{1}{\sqrt{5}} \left( \,
2|\uparrow\rangle^1 |0\rangle_x^1 |\alpha\rangle_x^2 +
|\downarrow\rangle^1 |\alpha\rangle_x^1 |0\rangle_x^2 \, \right)
\, |\uparrow\rangle^2 \, |\uparrow\rangle^3 |0\rangle_x^3 \, .
\end{equation}
A suitable pulse is then applied by lasers B2 and C2 to atom 2,
causing the internal state transformation
\begin{equation}
|\uparrow\rangle^2 \rightarrow \frac{1}{\sqrt{2}} \left( \,
|\uparrow\rangle^2 + |\downarrow\rangle^2 \, \right) \, .
\end{equation}
State transfer pulses (lasers A2 and A3) are now applied to atoms 
2 and 3, inducing the transformation
\begin{equation}
|\uparrow\rangle^2 |\alpha\rangle_x^2 \, |\uparrow\rangle^3
|0\rangle_x^3 \rightarrow 
|\uparrow\rangle^2 |0\rangle_x^2 \, |\uparrow\rangle^3
|\alpha\rangle_x^3 \, ,
\end{equation}
and producing an overall state
\begin{eqnarray}
\frac{1}{\sqrt{10}} && \left( \, 
2|\uparrow\rangle^1 |\uparrow\rangle^2 |0\rangle_x^1 
|0\rangle_x^2 |\alpha\rangle_x^3 \right. \nonumber
\\
&& \;\;\; + \;
2|\uparrow\rangle^1 |\downarrow\rangle^2 |0\rangle_x^1 
|\alpha\rangle_x^2 |0\rangle_x^3  \nonumber
\\
&& \;\;\; + \;
|\downarrow\rangle^1 |\uparrow\rangle^2 |\alpha\rangle_x^1 
|0\rangle_x^2 |0\rangle_x^3 \nonumber
\\
&& \;\;\; \left. + \;
|\downarrow\rangle^1 |\downarrow\rangle^2 |\alpha\rangle_x^1 
|0\rangle_x^2 |0\rangle_x^3 \, \right)
\, |\uparrow\rangle^3 \, .
\end{eqnarray}
Lasers B1, C1 and B2, C2 are finally applied to atoms 1 and 2 to 
generate the internal state transformations
\begin{eqnarray}
|\uparrow\rangle^j &\rightarrow & \frac{1}{\sqrt{2}} \left( \,
|\uparrow\rangle^j - |\downarrow\rangle^j \, \right) \, ,
\nonumber
\\
|\downarrow\rangle^j &\rightarrow & \frac{1}{\sqrt{2}} \left( \,
|\uparrow\rangle^j + |\downarrow\rangle^j \, \right) \, ,
\end{eqnarray}
with $j=1,2$, 
after which the overall state becomes
\begin{eqnarray}
&& \frac{1}{\sqrt{10}}\, |\uparrow\rangle^1 |\uparrow\rangle^2
\left( \, |0\rangle_x^1 |0\rangle_x^2 |\alpha\rangle_x^3
+ |0\rangle_x^1 |\alpha\rangle_x^2 |0\rangle_x^3 \right.
\nonumber
\\
&& \;\;\;\;\;\;\;\;\;\;\;\;\;\;\;\;\;\;\;\;\;\;\;\;\; \left.
+ \, |\alpha\rangle_x^1 |0\rangle_x^2 |0\rangle_x^3
\, \right) \, |\uparrow\rangle^3 \nonumber
\\
+ && \,\frac{1}{\sqrt{10}}\, |\downarrow\rangle^1 |\downarrow\rangle^2
\left( \, |0\rangle_x^1 |0\rangle_x^2 |\alpha\rangle_x^3
- |0\rangle_x^1 |\alpha\rangle_x^2 |0\rangle_x^3 \, \right)
\, |\uparrow\rangle^3 \nonumber
\\
+ && \,\frac{1}{\sqrt{10}}\, |\uparrow\rangle^1 |\downarrow\rangle^2
\left( \, -|0\rangle_x^1 |0\rangle_x^2 |\alpha\rangle_x^3
+ |0\rangle_x^1 |\alpha\rangle_x^2 |0\rangle_x^3 \, \right)
\, |\uparrow\rangle^3 \nonumber
\\
+ && \,\frac{1}{\sqrt{10}}\, |\downarrow\rangle^1 |\uparrow\rangle^2
\left( \, -|0\rangle_x^1 |0\rangle_x^2 |\alpha\rangle_x^3
- |0\rangle_x^1 |\alpha\rangle_x^2 |0\rangle_x^3 \right.
\nonumber
\\
&& \;\;\;\;\;\;\;\;\;\;\;\;\;\;\;\;\;\;\;\;\;\;\;\;\; \left.
+ \, |\alpha\rangle_x^1 |0\rangle_x^2 |0\rangle_x^3
\, \right) \, |\uparrow\rangle^3 \, .
\end{eqnarray}
Internal state detection of atom 2 in the level $|\uparrow\rangle^2$ 
(via laser D2) projects the system onto a state where,
once again, correlated with each internal state of atom 1 is
a delocalised motional coherent state, only now the state is 
delocalised between three atoms. A further internal state projection
onto $|\uparrow\rangle^1$ will of course disentangle internal and 
external degrees of freedom to yield the state
\begin{eqnarray}
\frac{1}{\sqrt{\cal N}} && \left( \, 
|0\rangle_x^1 |0\rangle_x^2 |\alpha\rangle_x^3 +
|0\rangle_x^1 |\alpha\rangle_x^2 |0\rangle_x^3 +
|\alpha\rangle_x^1 |0\rangle_x^2 |0\rangle_x^3 \, \right)
\nonumber
\\
&& \;\;\;\;\;\;\;\;\;\; \cdot \, |\uparrow\rangle^1 
|\uparrow\rangle^2 |\uparrow\rangle^3 \, .
\end{eqnarray}
Extensions of the scheme to more than three atoms follow 
straightforwardly, although the probability of obtaining the
desired sequence of projective measurement results obviously
decreases.

\section{
Other possibilities
}

The cases considered above clearly represent only a small subset
of the state manipulations that are in principle possible with the
trapped-atom-cavity configuration. 
Other combinations and sequences of internal 
state transformations and motional state transfers can produce a
wide variety of entangled and delocalised quantum states. While we
have not explored the possibilities exhaustively, two particular 
examples seem worthy of note. 

\subsection{
GHZ states
}

Consider again the state produced in the transformation of Eq.~(\ref{eq1}),
only now with an arbitrary motional state $|\phi\rangle_x^j$ replacing the
coherent state:
\begin{equation}
\frac{1}{\sqrt{2}} \left(\, |\uparrow\rangle^1 |0\rangle_x^1 
|\phi\rangle_x^2 + |\downarrow\rangle_1 |\phi\rangle_x^1 
|0\rangle_x^2 \,\right) \, |\uparrow\rangle^2 .
\end{equation}
This has the form of a Greenberger-Horne-Zeilinger state for three 
entangled degrees of freedom \cite{GHZ89}. One degree of freedom is
associated with the internal state of atom 1, while the other two
degrees of freedom correspond to the (external) motional states of
atoms 1 and 2 \cite{Wodkiewicz93Lange00}.

To physically separate the three degrees of freedom involved in the
GHZ state, one could implement the following strategy.
First, lasers B1 and C1 are applied to atom 1 to produce the internal
state transformation 
$\{\, |\uparrow\rangle^1\rightarrow -|\downarrow\rangle^1,\;
|\downarrow\rangle^1\rightarrow |\uparrow\rangle^1\,\}$.
Then, the output channel from cavity 1 is {\em redirected} to a {\em third}
atom-cavity system (so that atom 2 does not participate in the ensuing
state transfer), and appropriate state transfer pulses from lasers
A1 and A3 produce the transformation
\begin{eqnarray}
\frac{1}{\sqrt{2}} && \left(\, -|\downarrow\rangle^1 |0\rangle_x^1 
|\phi\rangle_x^2 + |\uparrow\rangle_1 |\phi\rangle_x^1 
|0\rangle_x^2 \,\right) \, |\uparrow\rangle^2 
|\uparrow\rangle^3 |0\rangle_x^3 \nonumber
\\
&& \rightarrow \frac{1}{\sqrt{2}}
\left(\, -|\downarrow\rangle^1 |\phi\rangle_x^2
|0\rangle_x^3 + |\uparrow\rangle_1 |0\rangle_x^2
|\phi\rangle_x^3 \,\right) \, 
|0\rangle_x^1 |\uparrow\rangle^2 |\uparrow\rangle^3 . \nonumber
\\
\end{eqnarray}
The entanglement is now between the internal state of atom 1 and
the motional states of atoms 2 and 3, with the three atoms possibly 
separated by large distances. Note that entanglement of the three
motional degrees of freedom could then be achieved by applying 
(via laser fields) an
internal-state-dependent motional transformation to atom 1; for
example, a transformation $\hat{U}_1(\phi )$ such that 
$\hat{U}_1(\phi )|\uparrow\rangle^1|0\rangle_x^1\rightarrow
|\uparrow\rangle^1|\phi\rangle_x^1$, while
$\hat{U}_1(\phi )|\downarrow\rangle^1|0\rangle_x^1\rightarrow
|\downarrow\rangle^1|0\rangle_x^1$.

\subsection{
Hardy state
}

Consider the situation in which the initial motional state of atom
1 is prepared as the simple superposition of Fock states
\begin{equation}
|\phi\rangle_x^1 = a|0\rangle_x^1 + b|1\rangle_x^1 \, , \;\;\;\;\;\;\;\;
(|a|^2+|b|^2=1) .
\end{equation}
Following the procedure outlined by Eqs.~(\ref{eq1}--\ref{2atom})
the (nonmaximally) entangled motional state that results is
\begin{equation}
\frac{1}{\sqrt{2(|a|^2+1)}} \, \left(\,
2a |0\rangle_x^1 |0\rangle_x^2 + b |0\rangle_x^1 |1\rangle_x^2
+ b |1\rangle_x^1 |0\rangle_x^2 \,\right) ,
\end{equation}
which is an example of what may be called a {\em Hardy state} 
\cite{Hardy93,Goldstein94,Franke00}. 
Note that, for the purpose of preparing such a state, 
$|1\rangle_x^1$ could be replaced by any Fock 
state $|n\rangle_x^1$ with $n>0$, or indeed by any 
(possibly mesoscopic) state orthogonal to $|0\rangle_x^1$.

\section{
Discussion
}

\subsection{
Possible applications
}

The delocalised states prepared by the present scheme evidently
offer a number of very interesting possibilities in a variety of
different contexts, particularly given the exquisite control with
which the states of trapped atoms can now be manipulated
(see, for example, \cite{Wineland98a,Wineland98b,Roos99}), and the
flexibility and efficiency that is in principle possible with 
regards to measurements (see, for example, 
\cite{Wineland98b,Leibfried96,Gardiner97}).
It is also worth emphasising again the ability to transfer the states of 
the motional modes onto propagating light fields with well-defined 
spatial and temporal properties [Eq.(\ref{aout})]. The resulting 
delocalised light fields could of course be subjected to standard 
optical manipulations and measurements.

Some possible applications are to:

\begin{enumerate}
\item
{\em Tests of quantum mechanics versus local realism}:\\
The GHZ and Hardy states just discussed allow for measurement outcomes
from a {\em single} set of observations that are forbidden by theories 
based on local realism (`nonlocality without inequalities').
Entangled coherent states of the form
${\cal N}^{-1/2}(|0\rangle_x^1|\alpha\rangle_x^2+|\alpha\rangle_x^1
|0\rangle_x^2)$ also admit tests of local realism, but using 
Bell-type inequalities \cite{Sanders92,Mann95,Rice00}, while also 
offering new perspectives on complementarity \cite{Rice00}.
In the context of tests of nonlocality, 
the fact that we are dealing with the states of {\em massive particles}
is potentially also a very significant feature with regards to issues of 
causality \cite{Fry98}.

\item
{\em Quantum networks and error correction in quantum computation}:\\
The possibility of establishing entanglement and transferring 
quantum information between different nodes of a quantum network
is naturally of great interest to the fields of quantum communication
and quantum computing \cite{Bouwmeester00}.
Indeed, quantum computer processors which are based on combinations of 
trapped-atom and cavity-QED methods and which communicate 
with other processors by optical means 
(via the cavity input-output coupling) have been proposed
(see, for example, \cite{Steane00}) 
and are amongst the leading contenders for 
initial demonstrations of substantial quantum computations.
States of the form $2^{-1/2}(|n\rangle^1|0\rangle^2+|0\rangle^1
|n\rangle^2)$ (with $n\geq 2$), where $|n\rangle^j$ is a
Fock state,
are also relevant to the idea of 
reversible measurements on a quantum system, or `quantum jump 
inversion' \cite{Mabuchi96}, which is in turn of relevance
to quantum error correction in quantum computing 
\cite{Steinbach00}.

\item
{\em Investigations of perceptual processes}:\\
If, for example, having prepared a state of the form 
$2^{-1/2}(|10\rangle_x^1|0\rangle_x^2+|0\rangle_x^1|10\rangle_x^2)$, 
one switches on lasers A1 and A2 and separates the two 
cavity output fields in space, then the result will be 
a 10-photon light pulse delocalised between two distinct
paths \cite{Gheri97}.
If these light pulses should be directed from distinct regions of
space towards the retina of a single (possibly human) observer, 
then Ghirardi \cite{Ghirardi99}
has suggested the possibility of new investigations into the linearity
(or nonlinearity) of quantum mechanics and the formation of definite 
perceptions.

\item
{\em Limits in interferometry}:\\
The state $2^{-1/2}(|n\rangle^1|0\rangle^2+|0\rangle^1|n\rangle^2)$
is known to allow the optimum $1/n$ phase sensitivity in a 
two-mode interferometer \cite{Wineland98b} and so is 
potentially of interest in the context of precision measurements. 
Entangled light fields of this form (which could be generated 
in the manner described in the previous application) have also been 
proposed for use in interferometric optical lithography, where they 
allow one in principle to beat the diffraction limit and write 
features much smaller than the optical wavelength \cite{Boto00}.

\end{enumerate}

\subsection{
Practical issues
}

The validity of, and constraints set by, the assumptions made in
deriving the motion-to-light state exchange model that is central 
to the present work are discussed and examined in more detail elsewhere 
\cite{Parkins99,Parkins00}.
Briefly, some of the more significant of these assumptions are as follows.

\begin{enumerate}
\item
{\em Lamb-Dicke assumption}:\\
Taking into account the spread of the atomic wave function with
increasing mean excitation number $\bar{n}_x$, 
the approximation $\sin (k\hat{x})\simeq\eta_x(\hat{b}_x+
\hat{b}_x^\dagger )$ can be associated with a condition of the form
$\eta_x^2(1+\bar{n}_x+3\sigma_{\bar{n}_x})/2\ll 1$, where 
$\sigma_{\bar{n}_x}^2$ is the variance of the number state distribution
\cite{Parkins00}. If we consider, for example, a coherent state of mean 
excitation number $\bar{n}_x=|\alpha |^2=10$, this condition reduces to
$\eta_x^2\ll 0.1$. 

\item
{\em Trap frequency and cavity linewidth}:\\
Simulations of the atom-cavity configuration and the state transfer
procedure show that the rotating-wave
approximation (with respect to the trap frequency) requires that 
$\nu_x$ be at least five to ten times
larger than the cavity field decay rate $\kappa$ 
\cite{Parkins99,Parkins00}.

\item
{\em Atomic spontaneous emission}:\\
Finite population in the atomic excited state leads to a finite 
degree of atomic spontaneous emission. For the effects of spontaneous
emission on the atomic motional state to be negligible for the duration
of the state transfer process (which occurs on a timescale of the order
of $\Gamma^{-1}$), the condition $g_0^2/(\kappa\gamma )\gg 1$ is
desirable \cite{Parkins99}. This is simply the condition of
strong coupling in cavity QED. 

\item
{\em Propagation losses}:\\
Throughout this work we have of course assumed ideal (i.e., lossless)
propagation of the light fields from one cavity to another. Photon 
losses will clearly degrade the fidelity of the state transmission;
some investigation along these lines is presented in \cite{Parkins00},
where, in particular, reflection of photons from the second cavity is 
allowed for. 
It is also noted, however, that photons `lost' from the system in this
way can in principle be detected and so postselection could be used 
to isolate `successful' transmissions.

\end{enumerate}

\noindent
Satisfying all of these conditions and constraints obviously 
constitutes a tremendous experimental challenge, although there is
reason to be hopeful.
Lamb-Dicke parameters of the order of 0.1 or smaller, and trapping
frequencies of several MHz to tens of MHz are now routinely achieved
in ion-trapping experiments 
(see, for example, \cite{Wineland98a,Roos99}). Optical dipole traps or
magnetic traps are also capable of confining neutral atoms in the 
Lamb-Dicke regime with frequencies in the MHz range
(see, for example, \cite{Ye99,Fortagh00}).

Also encouraging are the first generation of experiments in which single 
(neutral) atoms are trapped inside optical cavities in the regime of 
strong coupling cavity QED \cite{Ye99,Hood00,Pinkse00}.
These experiments were not in a regime where the effective
trapping frequency is larger than $\kappa$ (although effective 
Lamb-Dicke parameters along the cavity axis were small), 
but given the aforementioned possibilities for trapping and given 
probable further improvements in mirror technology (that is, higher 
cavity finesses), it seems reasonable to expect that future experiments 
trapping single atoms or ions inside optical cavities will be
able to meet the various criteria of our scheme simultaneously. 

With regards to intercavity propagation of possibly nonclassical
light fields, one can point, for example, to the experiment of
Turchette {\em et al}. \cite{Turchette98b}, in which squeezed light
from an optical parametric oscillator was transported through free 
space over a distance of several metres to a cavity-QED experimental
configuration. Transmission and mode-matching efficiencies in excess
of 90\% were achieved.

An estimate for likely timescales involved in the present scheme
can be given; if, for example, one assumes a trap frequency of 
$\nu_x/(2\pi )\simeq 5\,{\rm MHz}$ and a cavity field decay rate of
$\kappa /(2\pi )\simeq 500\,{\rm kHz}$, then an estimate for
the rate of state transfer might be 
$\Gamma /(2\pi )\simeq 5-20\,{\rm kHz}$, or 
$\Gamma^{-1}\simeq 10-30\,\mu {\rm s}$. 
Note that the other basic operations involved in the scheme using the 
auxiliary lasers B,C, and D are routinely performed with very
high efficiency in trapped-ion experiments, while the timescale 
for decoherence of motional states of single trapped ions is typically 
of the order of milliseconds or even longer
\cite{Wineland98a,Roos99,Turchette00}. This intrinsically slow
decoherence rate of the motional state would allow for (relatively)
long-lived delocalised quantum superpositions, which is an important
and attractive feature of the proposal put forward in this work.

\acknowledgments
The author thanks H.J. Kimble, H. Ritsch, and D. Leibfried for 
discussions and gratefully acknowledges support from the Marsden Fund 
of the Royal Society of New Zealand. He also thanks the Quantum Optics
groups at the University of Innsbruck and the California Institute
of Technology for support and hospitality during visits when part 
of this work was carried out.

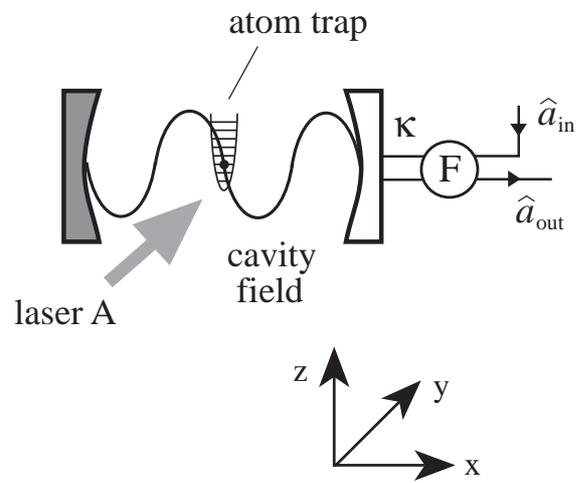
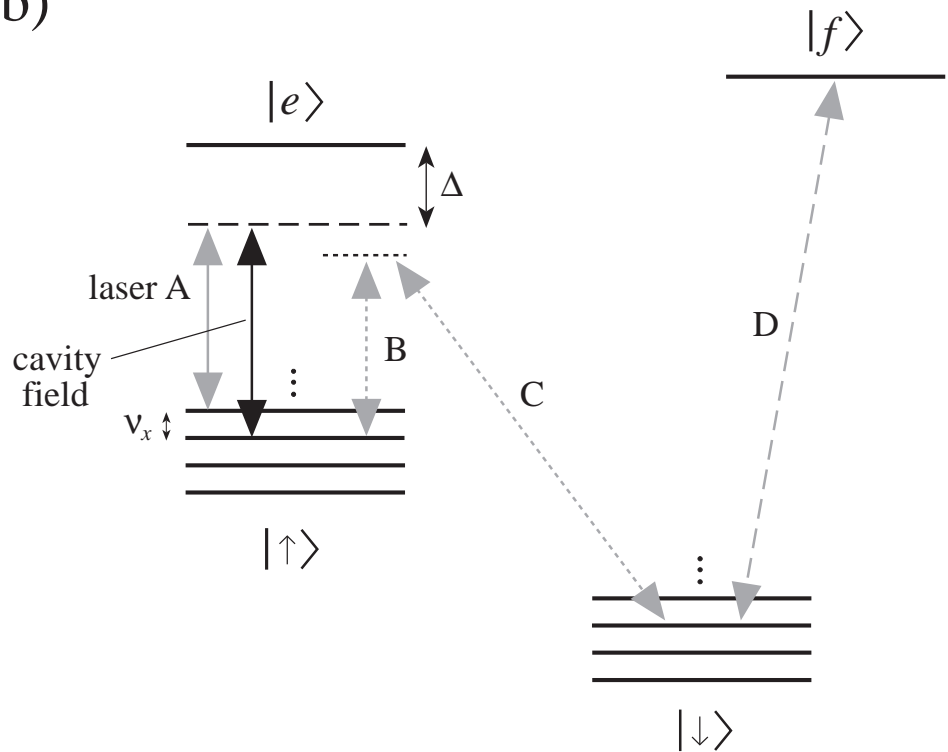
\begin{figure}
\caption{
Schematic of proposed (a) setup and (b) excitation scheme 
for quantum state exchange between the motion of a trapped atom 
and a quantised cavity mode of the electromagnetic field.
Note that all input and output to the atom-cavity system is 
through just one mirror; the other mirror is assumed to be perfect.
The internal atomic structure shown in (b) is like that 
occurring in recent ion trap experiments. For simplicity, the
vibrational level structure is shown only for the internal levels
$|\uparrow\rangle$ and $|\downarrow\rangle$.
}
\end{figure}

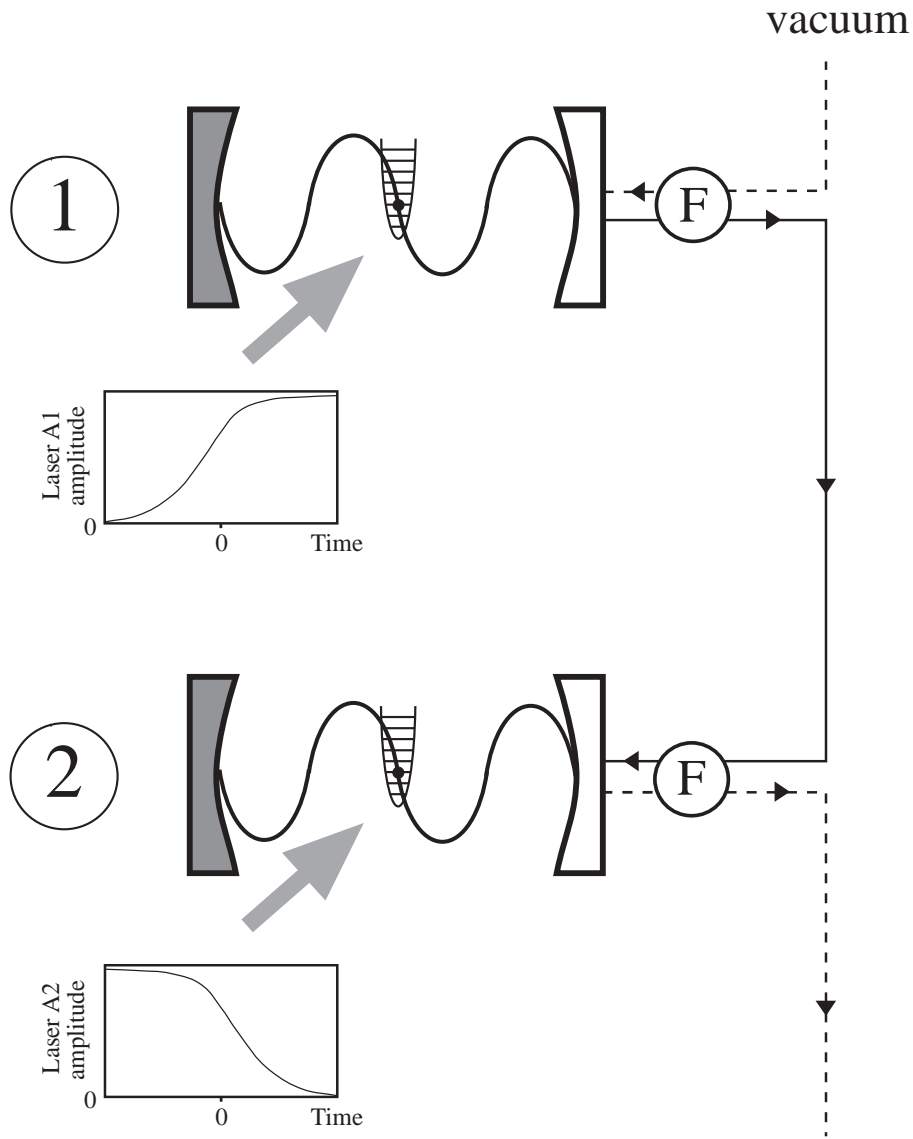
\begin{figure}
\caption{
Configuration for quantum transmission of the motional state 
of a trapped atom from site 1 to site 2. Lasers A1 and A2
are applied with time-dependent amplitudes of the forms shown,
while Faraday isolators (F) facilitate a separation of input
and output channels to and from the atom-cavity systems.
Note that, during an ideal transfer, no light escapes from 
cavity 2 through its output channel. 
}
\end{figure}


\begin{thebibliography}{99}
\bibitem{PulseShapes}
Note that, as shown in \cite{Parkins99}, it is also possible to
to design pulse shapes to produce the state transfer 
(\ref{statetransfer}) in the case where the cavity mode dynamics
is retained in the model; that is, when the cavity modes are not 
adiabatically eliminated from the system dynamics.
\end{thebibliography}
\end{document}